\documentclass[aps,prl,twocolumn,groupedaddress,showpacs]{revtex4}

\usepackage{graphicx}
\begin{document}

\title{Quantum Tunneling of Magnetization in a Large Molecular 
Nanomagnet -- Approaching the Mesoscale}

\author{Wolfgang Wernsdorfer$^1$, Monica Soler$^{2}$,  
David N. Hendrickson$^3$, George Christou$^{2}$}

\affiliation{
$^1$Lab. L. N\'eel, associ\'e \`a l'UJF, CNRS, BP 166,
38042 Grenoble Cedex 9, France\\
$^2$Dept. of Chemistry, Univ. of Florida, 
Gainesville, Florida 32611-7200, USA\\
$^3$Dept. of Chemistry, Univ. of California at San 
Diego, La Jolla, California 92093-0358, USA
}

\date{\today}

\begin{abstract}
A Mn$_{30}$ molecular cluster is established to be the 
largest single-molecule magnet (SMM) discovered to date.
Magnetization versus field measurements show coercive fields of
about 0.5 T at low temperatures.
Magnetization decay experiments reveal an Arrhenius behavior 
and temperature-independent 
relaxation below 0.2 K diagnostic of quantum tunneling of 
magnetization through the anisotropy barrier.
The quantum hole digging method is used to establish
resonant quantum tunneling. 
These results demonstrate that large molecular nanomagnets,
having a volume of 15 nm$^3$,
with dimensions approaching the mesoscale can still exhibit 
the quantum physics of the microscale.
\end{abstract}


\maketitle

The study of the interface between classical 
and quantum physics has always been a 
fascinating area, but its importance has 
nevertheless grown dramatically with the current 
explosive thrusts in nanoscience. Taking devices 
to the limit of miniaturization 
(the mesoscale and beyond) where quantum effects 
become important makes it essential to understand 
the interplay between the classical properties 
of the macroscale and the quantum properties of 
the microscale. This is particularly true 
in nanomagnetism, where many potential 
applications require monodisperse, magnetic 
nanoparticles. One source of such species are 
single-molecule magnets (SMMs)
~\cite{Christou00,Sessoli93b,Sessoli93,Aubin96,Boskovic02}, 
individual molecules that function as 
single-domain magnetic particles. 
Below their blocking temperature, they exhibit 
magnetization hysteresis, the classical 
macroscale property of a magnet, as well as 
quantum tunneling of magnetization (QTM)
~\cite{Friedman96,Thomas96,Sangregorio97,Aubin98,Aubin98b,Boskovic01} 
and quantum phase interference~\cite{WW_Science99,Garg93}, the properties 
of a microscale entity. QTM is advantageous 
for some potential applications of SMMs, 
e.g. in providing the quantum superposition 
of states for quantum computing~\cite{Leuenberger01}, 
but it is a disadvantage in others such as 
information storage where it would lead to 
loss of preferential spin alignment. 
Large SMMs approaching the mesoscale have 
long been the target of synthesis, and an 
important question for these (and other 
mesoscale magnetic particles) is whether they 
might still unequivocally demonstrate 
quantum properties, as theoretically predicted~\cite{Chudnovsky88,Barbara90}. 

In this letter, we show that a Mn$_{30}$ molecular cluster
is by far the largest SMM to date, and show 
that it unambiguously still exhibits QTM. 
This establishes that the quantum physics 
of the microscale can also be observed in 
large molecular nanomagnets.  

\begin{figure}
\begin{center}
\includegraphics[width=.45\textwidth]{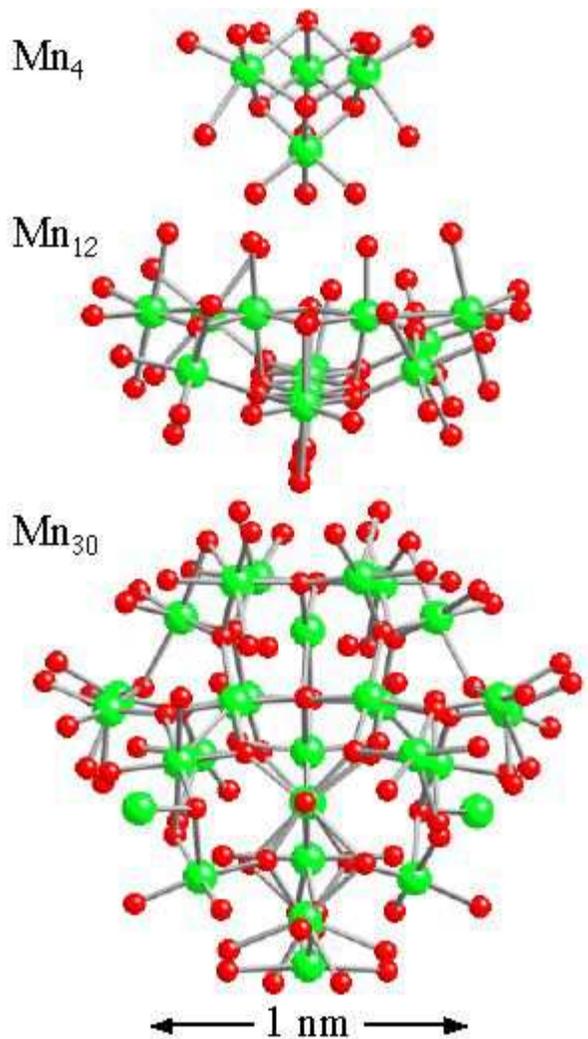}
\caption{Schematic view of the Mn/O magnetic core
of a Mn$_4$~\cite{Aubin96}, Mn$_{12}$~\cite{Boskovic02}, and Mn$_{30}$~\cite{Soler01}
SMMs. The Mn atoms are depicted larger than the O atoms.}
\label{fig1}
\end{center}
\end{figure}

The Mn$_{30}$ molecular cluster has the formula
[Mn$_{30}$O$_{24}$(OH)$_8$(O$_2$CCH$_2$CMe$_3$)$_{32}$(H$_2$O)$_2$(MeNO$_2$)$_4$]. xCH$_2$Cl$_2$.yH$_2$O 
and crystallizes in the monoclinic space group C$_{\rm 2/c}$ 
with each Mn$_{30}$ on a crystallographic C$_{\rm 2}$ 
rotation axis~\cite{Soler01}. The molecules are 
thus all parallel with their molecular C$_{\rm 2}$ 
axes ($z$ axes) aligned along the crystal $b$ axis. 
Each Mn$_{30}$ comprises a Mn/O magnetic core (Fig. 1) 
of volume $\approx$4 nm$^3$ enveloped in 
a hydrocarbon shell $\approx$0.5 nm thick 
composed of carboxylate CH$_2$CMe$_3$ groups. 
This essentially magnetically isolates neighboring Mn/O cores. 
However, the large size of Mn$_{30}$ 
(volume $\approx$15 nm$^3$, molar mass $\approx$6335 Daltons), 
akin to a small protein, causes inefficient 
packing in the crystal, with voids between molecules 
that contain disordered solvent molecules, primarily CH$_2$Cl$_2$. 
In addition, some CMe$_3$ groups are statically 
disordered about two sites.

Each Mn$_{30}$ contains one Mn$^{4+}$, twenty-six Mn$^{3+}$ 
and three Mn$^{2+}$ ions, and exchange coupling between 
the many spins of the Mn/O core leads to Mn$_{30}$ 
having a small net uncompensated spin 
in the ground state of S = 5.  
The latter was determined from temperature 
and field magnetization measurements on powdered, 
microcrystalline samples in the 1.8-4.0 K and 0.1-0.4 T ranges. 
The data were fit by diagonalization of the spin Hamiltonian 
matrix assuming only the ground state is populated and 
incorporating axial anisotropy $(D S_z^2)$ 
and Zeeman terms, and employing a full powder average~\cite{Yoo01}; 
the Mn30 is thus modelled as a Ôgiant spinÕ 
with Ising-like anisotropy.  
The corresponding Hamiltonian is given by equation (1):
\begin{equation}
	\mathcal{H} = -D S_z^2 + \mathcal{H}_{{\rm trans}} 
	+ g \mu_{\rm B} \mu_0 \vec{S}\cdot\vec{H} 
\label{eq_H}
\end{equation}
where $D$ is the axial anisotropy constant,
$S_{z}$ is the $z$-component of the spin operator $\vec{S}$,
$\mathcal{H}_{{\rm trans}}$ is the transverse anisotropy, 
containing $S_x$ and $S_y$ operators, 
$g$ is the electronic g-factor, 
$\mu_{\rm B}$ is the Bohr magneton, 
$\mu_0$ is the vacuum permeability, 
and $\vec{H}$  is the applied field. 
The last term in eq. 1 is the Zeeman energy 
associated with an applied field. 
The fit parameters were $S = 5$, 
$D$ = -0.73 K, and $g = 2.00$, 
with the transverse anisotropy $\mathcal{H}_{{\rm trans}}$ not 
included for simplicity (in fact, inclusion of a 
small transverse anisotropy term involving the 
transverse anisotropy constant, $E$, had no significant 
effect on the fit). These values suggest an 
upper limit to the potential energy barrier $U$ 
to magnetization relaxation (reversal) of $U = S^2|D| = 18$ K, 
a value potentially large enough to make Mn$_{30}$ a SMM 
at sufficiently low temperatures. 
This was therefore explored by hysteresis loop,
relaxation, and hole digging measurements.

\begin{figure}
\begin{center}
\includegraphics[width=.45\textwidth]{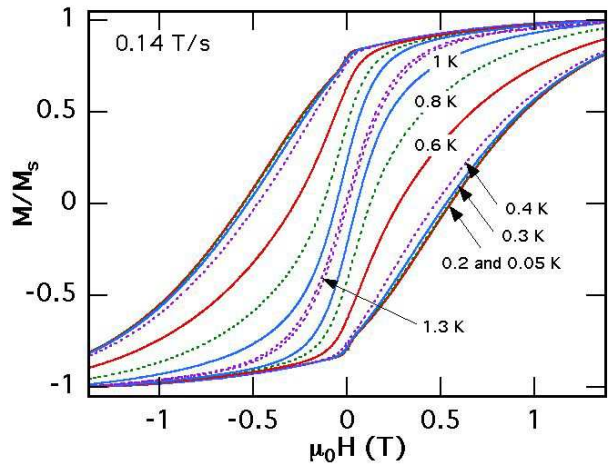}
\caption{Normalized magnetization of Mn$_{30}$ versus 
applied magnetic field in the 0.05 to 1.0 K temperature range  
and a field sweep rates of 0.14 T/s. 
Note that the loops become temperature-independent 
below about 0.2 K.}
\label{fig2}
\end{center}
\end{figure}

Studies were performed on single crystals using 
an array of micro-SQUIDs~\cite{WW_ACP_01}. 
Hysteresis in magnetization versus magnetic field 
scans was observed, establishing Mn$_{30}$ as a new SMM, 
the largest to date. Fig. 2 shows typical 
hysteresis loops, with the field applied approximately 
along the easy axis of magnetization. 
The blocking temperature is about 1.2 K, and the 
coercivity increases with decreasing temperature 
and increasing sweep rate, as expected for the 
superparamagnet-like behavior of a SMM. 
The loops do not display step-like features 
characteristic of resonant QTM between the energy 
states of the Mn$_{30}$ molecule, as observed 
for several other SMMs
~\cite{Friedman96,Thomas96,Sangregorio97,Aubin98,Aubin98b,Boskovic01,WW_Science99}. 
However, it is possible 
that steps are present but smeared out by broadening 
effects due to a distribution of magnetization 
relaxation barriers (i.e. $D$ values), consistent 
with the distribution of Mn$_{30}$ environments 
resulting from the disordered CMe$_3$ groups and 
solvent molecules observed in the crystal structure. 
It is thus not possible to determine directly 
from the hysteresis loops whether resonant QTM is occurring 
in this large Mn$_{30}$ molecule. 
Nevertheless, the first indication of quantum tunneling is the 
temperature independence of the hysteresis loops
below about 0.3 K (Fig.2). Detailed measurements
of the hysteresis loops show however that the
loops are still time-dependent below 0.3 K.
This is presented in the inset of Fig. 3 showing the temperature
dependence of the coercive field at different
field sweep rates. The strong time-dependence in the
temperature-independent regime is another important
indication of QTM.

\begin{figure}
\begin{center}
\includegraphics[width=.45\textwidth]{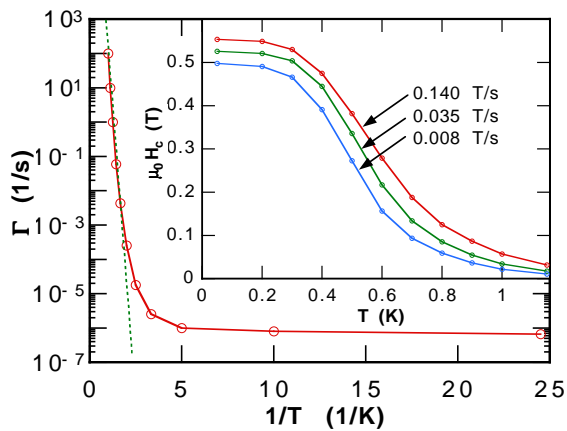}
\caption{Arrhenius plot of the relaxation rate $\Gamma$ 
versus the inverse temperature for Mn$_{30}$ using 
data obtained from magnetization decay measurements. 
The dashed line is the fit of the data in the 
thermally activated region to the Arrhenius equation; 
see the text for the fit parameters. 
Inset: coercive field versus temperature for three field sweep 
rates.}
\label{fig3}
\end{center}
\end{figure}

QTM can also be confirmed with 
magnetization decay studies.  
The magnetization was first saturated at 5 K with 
a large applied field, the temperature decreased to 
a chosen value, and then the field removed 
and the magnetization decay monitored with time. 
This provided magnetization relaxation
rates $\Gamma$ at different temperatures
allowing us to make an Arrhenius plot (Fig. 3)
that is based on the Arrhenius law: 
\begin{equation}
	\Gamma = \Gamma_0 exp(-U_{\rm eff}/k_{\rm B}T) 
\label{tau}
\end{equation}
where $\Gamma_0$ is the pre-exponential factor, 
$U_{\rm eff}$ is the mean effective barrier 
to relaxation, and $k_{\rm B}$ is the Boltzmann 
constant. The slope in the thermally activated 
region above 0.5 K gave $U_{\rm eff}/k_{\rm B}$ = 15 K 
and $\Gamma_0 = 1.5 \times 10^{-8}$~s$^{-1}$. 
The mean barrier $U_{\rm eff}$ 
is smaller than the calculated $U = S^2|D| = 18$ K, 
as expected for QTM between higher energy levels $M_s$ 
of the $S = 5$ manifold. More importantly, the relaxation 
time becomes temperature-independent below ~0.2 K 
at about 10$^{-6}$ s$^{-1}$, diagnostic of ground state 
QTM between the lowest energy $m = \pm5$ levels. 
The crossover temperature between thermally activated 
relaxation and ground state tunneling is between 0.2 and 0.3 K. 
The ground state tunneling at $<$ 0.2 K rationalizes 
the appearance at these temperatures of a QTM step 
at H = 0 in the loops of Fig. 2. Since the tunneling 
is now only between the lowest $m$ levels, 
the distribution of Mn$_{30}$ environments no longer 
has such a large broadening effect on the step 
at zero field, and a better defined step is seen.

\begin{figure}
\begin{center}
\includegraphics[width=.45\textwidth]{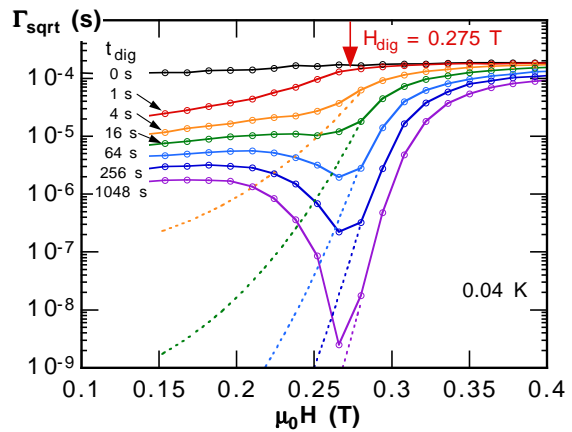}
\caption{Relaxation rate at short time periods $\Gamma_{\rm sqrt}$
versus applied field, determined after application 
for digging-time $t_{\rm dig}$ of a digging-field $H_{\rm dig}$ of 0.275T. 
The sharp dip ('hole') at the $H_{\rm dig}$ value shows 
the depletion of spins that were in resonance 
at $H_{\rm dig}$ and thus underwent resonant tunneling 
during $t_{\rm dig}$. The half-width of the hole 
of about 20 mT is larger than the hyperfine 
field (about 10 mT), reflecting the influence of dipolar couplings.
A classical system or a quantum system with an continuum 
of levels would not show a hole but a monotonic
increase for $H < H_{\rm dig}$, indicated schematically 
by the dotted lines.}
\label{fig4}
\end{center}
\end{figure}

Additional confirmation of QTM in Mn$_{30}$ 
was obtained from the `quantum hole-digging' method
~\cite{Prokofev98,WW_PRL99,WW_PRL00,Fernandez01,Tupitsyn03}.
This is a relatively new method that can, 
among other things, establish whether resonant tunneling 
occurs even when steps are absent in hysteresis 
loops due to a distribution of energy barriers. 
The method is based on the simple idea that 
after a rapid field change, the resulting 
magnetization relaxation at short time periods 
is directly related to the number of molecules 
in resonance at the applied field; 
Prokof'ev and Stamp proposed~\cite{Prokofev98} that this 
short time relaxation should follow a $\sqrt{t}$ 
($t$ = time) relaxation law. Thus, the magnetization 
of the Mn$_{30}$ molecules in the crystal was first 
saturated with a large negative field, and then a 
`digging field' $H_{\rm dig}$ of 0.275 T was 
applied at 0.04 K for a chosen `digging time' $t_{\rm dig}$. 
If QTM can occur in Mn$_{30}$, then that fraction 
(and only that fraction) of the molecules that 
is in resonance at $H_{\rm dig}$ can undergo 
magnetization tunneling. After $t_{\rm dig}$, a field 
$H_{\rm probe}$ is applied and the magnetization 
relaxation rate is measured for short time periods; 
from this is calculated the short-time relaxation 
rate $\Gamma_{\rm sqrt}$, which is related to the number 
of Mn$_{30}$ molecules still available for QTM~\cite{WW_ACP_01}. 
The entire procedure is then repeated at 
other $H_{\rm probe}$ fields. 
The resulting plot of $\Gamma_{\rm sqrt}$ versus $H_{\rm probe}$ 
reflects the distribution of spins still available 
for tunneling after $t_{\rm dig}$, and it will display 
a `hole' if resonant QTM did indeed occur during application 
of $H_{\rm dig}$ for time $t_{\rm dig}$.  
The obtained plot for Mn$_{30}$ (Fig. 4) does show a hole 
in this distribution, corresponding to a depletion of 
those spins that were in resonance and able to tunnel 
during $t_{\rm dig}$ at $H_{\rm dig}$. 
The occurrence of resonant QTM in Mn$_{30}$ is thus confirmed, 
since a classical system or a quantum system with
an continuum of levels will not show a `quantum hole'
but rather a depletion of all spins 
with low barriers (Fig. 4).

The above results unambiguously demonstrate 
that Mn$_{30}$ undergoes tunneling, involving the 
coherent reversal of about 600 interacting electron 
spins within the Mn/O magnetic core. Of the three diagnostic 
tests for resonant QTM, Mn$_{30}$ clearly 
demonstrates two of them, temperature-independent 
relaxation and quantum hole digging. The third is steps 
in hysteresis loops, but these are broadened beyond 
resolution by the distribution of barriers ($D$ values) 
resulting from a distribution of Mn$_{30}$ environments. 
Mn$_{30}$ thus represents the largest SMM by far 
to unequivocally demonstrate QTM, establishing that 
quantum effects can be clearly observed, and 
studied, even in very large 'magnetic particles' 
with dimensions approaching the mesoscale. 
This is made possible in Mn$_{30}$ by its monodisperse 
nature and highly ordered arrangement within the crystal 
(notwithstanding the solvent and CMe$_3$ disorder, 
which in an absolute sense is a small perturbation), 
which prevent complications from distributions 
of particle size, shape, surface roughness, and spin. 
The latter have severely hampered previous attempts 
to demonstrate QTM in large magnetic particles, 
both for classical magnetic materials such as Co metal 
and molecules such as the ferritin protein. The results 
have consequently usually been negative, unclear or 
controversial~\cite{Awschalom92,Gider95,Tejada96,Mamiya02,WW_PRL97_BaFeO}. 
In contrast, the present work 
demonstrates that crystalline assemblies of monodisperse 
molecular nanomagnets allow clear observation of 
quantum properties under straightforward conditions 
even for large systems approaching the mesoscale, 
the hydrocarbon shell of organic groups in Mn$_{30}$ 
ensuring the Mn/O cores are exactly identical in every molecule.  

From a broader point of view, 
the present work confirms that large SMMs 
can be prepared and studied as crystalline 
assemblies of monodisperse particles. 
Mn$_{30}$ is much bigger than other currently 
known SMMs such as Mn$_{4}$, Mn$_{12}$, etc, and thus 
represents proof-of-feasibility of extending to larger 
species the many advantages of SMMs over classical magnetic 
particles. These include monodispersity, crystalline order, 
room-temperature solution synthesis, true solubility 
(rather than colloidal suspension) in common organic 
solvents, and an insulating shell of organic groups 
around the magnetic core, which additionally can 
be varied at will using standard synthetic 
chemistry methods. Such factors are clear benefits 
for the many proposed applications of SMMs, 
for example as qubits in quantum computing~\cite{Leuenberger01},
where the larger is an individual SMM, the easier it 
should be to manipulate and measure its quantum state. 
The synthetic variation of the hydrocarbon shell can 
also provide a facile way to increase (or decrease, 
if desired~\cite{WW_Nature02}) the isolation of a single 
SMM from its neighbors, and perhaps thus also 
help decrease decoherence problems.  

We thank the USA National Science Foundation 
for support.

%\bibliographystyle{wernsdor} 
%\bibliography{wernsdor}

\end{document}